# Identifying Optimal Methods for Addressing Confounding Bias When Estimating the Effects of State-Level Policies

Beth Ann Griffin,[1] Megan S. Schuler,[1] Elizabeth M. Stone,[2] Stephen W. Patrick,[3,4] Bradley D. Stein,[4] Pedro Nascimento de Lima,[1] Max Griswold,[1] Adam Scherling,[5] Elizabeth A. Stuart[2]

[1]RAND Corporation, Arlington, VA
[2] Johns Hopkins Bloomberg School of Public Health, Baltimore, MD, USA
[3]Department of Pediatrics, Vanderbilt University, Nashville, Tennessee
Mildred Stahlman Division of Neonatology, Vanderbilt University, Nashville, Tennessee
Vanderbilt Center for Child Health Policy, Nashville, Tennessee
Department of Health Policy, Vanderbilt University, Nashville, Tennessee
[4]RAND Corporation, Pittsburgh, Pennsylvania
[5]Disney Streaming, Los Angeles, California

**Corresponding Author:**
Beth Ann Griffin
1200 South Hayes Street
Arlington, VA 22202
703-413-1100
 bethg@rand.org

**Running title:** Optimal methods for addressing confounding bias

**Conflicts of Interest:** The authors declare that they have no competing interests.

**Sources of Funding:** This research was financially supported through a National Institutes of Health (NIH) grant (P50DA046351) to RAND (PI: Stein). NIH had no role in the design of the study, analysis, and interpretation of data nor in writing the manuscript.

**Data:** The data that support the findings of this study are available from National Vital Statistics System (NVSS) Multiple Cause of Death mortality files (1999 through 2016) but restrictions apply to the availability of these data, which were used under license for the current study, and so are not publicly available. The data can be request under a similar license (data use agreement) from the NVSS. The code and R package needed to run the simulations on any repeated measures data is available at: http://github.com/optic-tools/optic.

**Acknowledgments:** The authors would like to thank members of the RAND-USC Opioid Tools and Information Center (OPTIC) including Rosalie Pacula, Rosanna Smart, and David Powell as well as members of the RAND statistics group who provided helpful comments on this work as it progressed. Finally, the authors want to thank Hilary Peterson for her assistance with manuscript preparation and Pedro Nascimento de Lima for help with the graphics.

# Identifying Optimal Methods for Addressing Confounding Bias When Estimating the Effects of State-Level Policies

**Abstract**

Background: Policy evaluation studies that assess how state-level policies affect health-related outcomes are foundational to health and social policy research. The relative ability of newer analytic methods to address confounding, a key source of bias in observational studies, has not been closely examined.

Methods: We conducted a simulation study to examine how differing magnitudes of confounding affected the performance of four methods used for policy evaluations: (1) the two-way fixed effects (TWFE) difference-in-differences (DID) model; (2) a one-period lagged autoregressive (AR) model; (3) augmented synthetic control method (ASCM); and (4) the doubly robust DID approach with multiple time periods from Callaway-Sant'Anna (CSA). We simulated our data to have staggered policy adoption and multiple confounding scenarios (i.e., varying the magnitude and nature of confounding relationships).

Results: Bias increased for each method: (1) as confounding magnitude increases; (2) when confounding is generated with respect to prior outcome trends (rather than levels), and (3) when confounding associations are nonlinear (rather than linear). The AR and ASCM have notably lower root mean squared error than the TWFE model and CSA approach for all scenarios; the exception is nonlinear confounding by prior trends, where CSA excels. Coverage rates are unreasonably high for ASCM (e.g., 100%), reflecting large model-based standard errors and wide confidence intervals in practice.

Conclusions: Our simulation study indicated that no single method consistently outperforms the others. But a researcher's toolkit should include all methodological options. Our simulations and associated R package can help researchers choose the most appropriate approach for their data.

**Key Words:** Confounding, State-level policy, Policy evaluations, Difference-in-differences, Synthetic Control Method, Autoregressive model, Opioid Overdose, Simulation

**INTRODUCTION**

Studies that assess how a given policy affects certain health-related outcomes are foundational to health and social policy research.[1,2] Evaluations of state-level policies face specific methodological challenges.[3] Confounding is a well-recognized source of bias in observational studies;[3-7] however, the unique considerations of confounding in longitudinal settings – including difference-in-differences (DID) evaluations – have received less attention.

A confounder is a third variable associated with both a treatment and outcome; however, this definition becomes more complex when these variables are time-varying. Confounding is more complex in DID than in standard comparison group designs because DID uses trends over time in the comparison groups as a proxy for what would have happened in the treated groups had they not adopted the policy. A recent simulation study characterizing confounding in a DID context established that confounders are covariates that evolve differently across time in the policy and comparison groups and have a time-varying effect on the outcome.[8] Importantly, to cause confounding in DID, a variable must be associated with outcome *trends* rather than simply outcome *levels*.[9] DID models control for confounding by including state- and time-fixed effects; time-varying confounders may also be included.

Failure to appropriately account for confounders can produce biased effect estimates in which longitudinal relationships between confounders and outcomes are not disentangled from true policy effects. In all contexts, controlling for post-treatment confounders (time-varying confounding variables impacted by the time-varying treatment) biases effect estimates.[10] Including time-varying confounders in a DID model may amount to controlling for post-treatment confounders when policy adoption dates vary across states.[8]

Applied policy evaluation studies have paid limited attention to confounding, perhaps because the key causal assumption in DID study designs is that outcome trends for treatment and control groups would have evolved similarly across time had the treatment group not been exposed to the policy.[11,12] DID designs do not require policy and comparison groups to be well matched on covariates; rather they simply require that the parallel counterfactual trend assumption is plausible.[8] Thus in practice, many DID studies do not identify and control for potential confounders.

Recent methodological work has focused on other methodological issues. Numerous studies have highlighted the challenges of using the common two-way fixed effects (TWFE) DID model to estimate policy effects in the presence of treatment effect heterogeneity. Recent methodological advances have sought to improve estimation of policy effects in the presence of staggered adoption, including use of autoregressive models (AR);[13] augmented synthetic control methods (ASCM);[14] and methods from Callaway Sant'Anna (CSA),[15] which directly allow for effect heterogeneity over time and by state. ASCM and CSA take a more design-based approach (rather than regression-based) to policy evaluation, which helps mitigate concerns about bias that stems from controlling for time-varying post-treatment confounding variables directly in the model.

We examined the role of confounding in longitudinal DID-type studies to understand the implications of confounding in state policy evaluations and the relative performance of traditional and recently developed methods. Using the opioid crisis as a motivating policy context, we conducted a simulation study to examine how differing magnitudes of confounding affected the performance of four methods (TWFE, AR, ASCM, and CSA) since these methods capture some of the more commonly used methods in this space and represent a range of the type of analytic approaches available to researchers for estimating the effects of state policies.

## 2. METHODS

In 2021, there were more than 80,000 fatal opioid-related overdoses,[16] motivating states to adopt multiple opioid-related policies.[17-19] Our simulated data comprise annual state-level data on the outcome (opioid-related mortality) and covariates; the data also reflect staggered policy adoption. The policy variable represents a hypothetical policy intended to reduce opioid-related mortality. The study was approved by the corresponding author's Institutional Review Board with a waiver of consent (Assurance Number: FWA00003425).

### 2.1 Inferential Goal

Using potential outcomes notation, let $Y_{it1}$ denote the potential outcome for state $i$ ($i = 1, \ldots, 50$) if the policy was in effect at time $t$, while $Y_{it0}$ denotes the potential outcome for state $i$ if the policy was not in effect at time $t$. Thus, each state has two potential outcomes at each time point, representing the mortality rates that would be observed with and without the policy in effect. Our primary treatment effect of interest is $E[Y_1 - Y_0]$, averaging across both states and times. Let $A_{it} = \{0,1\}$ denote an indicator for whether state $i$ had the policy in effect at time $t$ (where $t = 1, \ldots, T$). Then, $Y_{it}^{obs} = Y_{it1} * A_{it} + Y_{it0} * (1 - A_{it})$ denotes the observed outcome for state $i$ at time $t$ as measured longitudinally for state $i$ over time $t = 1, \ldots, T$. Let $X_{it}$ denote the vector of observed time-varying (or time-invariant) covariates at the state level.

### 2.2 Empirical Models Considered

We compared four statistical models: (1) classic TWFE model; (2) one-period lagged autoregressive (AR) model; (3) ASCM; and (4) doubly-robust DID approach with multiple time periods from CSA.[15] We chose these models based on previous policy simulation studies[6] and on recently proposed design-based methods for addressing confounding in policy evaluations.[20] The models differ in how they address confounding in longitudinal DID-type studies.

A commonly used model is the classic TWFE model. It aims to control for confounding using covariate adjustment and fixed effects.

DID compares the pre-policy to post-policy change in the treatment group (states that enacted the policy) to the corresponding pre-period to post-period change in the control group (non-enacting states). The classic DID specification is generally implemented as a two-way fixed effects model that includes both state- and time-fixed effects as well as key potential confounders captured in $X_{it}$, expressed as:

$$g(Y_{it}^{obs}) = \alpha \cdot A_{it} + \boldsymbol{\beta} \cdot \boldsymbol{X}_{it} + \rho_i + \sigma_t + \varepsilon_{it} \qquad (1)$$

where $g(.)$ denotes the generalized linear model link function (e.g., linear, log) and $\varepsilon_{it}$ denotes the error term. State-fixed effects, $\rho_i$, quantify differences in the outcome across states and time-fixed effects, $\sigma_t$, quantify national temporal trends. State-fixed effects account only for time-invariant differences between treated and untreated states; time-fixed effects account only for exogenous factors that affect both treated and untreated states equally. Additional covariates $X_{it}$ may be included to control for state-specific time-varying confounders.

The second model we considered was an AR model. AR models aim to minimize confounding by controlling for time-fixed effects, lagged outcome values, and other time-invariant or time-varying state-level confounders using $X_{it}$.

Previous simulation studies in gun policy and opioid policy contexts demonstrated that AR models perform especially well for estimating state-level policy effects.[6,13] AR models include one or more lagged measures of the outcome variable (e.g., $Y_{it-1}^{obs}$) as covariates to control for potential confounding from differences in prior outcome trends across treated and comparison states. AR models improve prediction when outcomes are highly autocorrelated, as with annual measures of state-level opioid-related mortality.

The AR model we examined—highly rated in prior simulation studies,[6,13] included a single lagged value of the outcome expressed as:

$$g(Y_{it}^{obs}) = \alpha \cdot (A_{it} - A_{i,t-1}) + \boldsymbol{\beta} \cdot \boldsymbol{X}_{it} + \gamma \cdot Y_{it-1}^{obs} + \sigma_t + \epsilon_{it} \tag{2}$$

Akin to Equation (1), this model includes time-fixed effects, $\sigma_t$, to quantify temporal trends across time but adjusts for state-specific variability through use of the AR term ($\gamma \cdot Y_{it-1}^{obs}$) rather than state fixed effects.

We also considered ASCM with staggered adoption.[21] This model aims to control for confounding in two different stages: first by using weighting to ensure comparability between treated and control states within each enactment year cohort and then via additional covariate adjustment in the outcome model.

As originally proposed, ASCM applies to situations in which there is only one treated unit (e.g., state) and several untreated units. Weights are assigned to the untreated units to create a weighted synthetic control group that most closely matches the treated unit on pre-treatment outcomes and potential confounders. The estimated treatment effect is calculated by comparing post-treatment outcomes between the treated unit and the synthetic control (i.e., a weighted average of the outcomes of the untreated units).

ASCM adds a bias-correction step that adjusts for any remaining pre-treatment imbalances in outcomes or covariates between the treated units and the synthetic control.[14] Expanding this approach to situations in which multiple units receive treatment, Ben-Michael et al.[14] proposed creating a separate augmented synthetic control for each treated unit and calculating the weights using a partial pooling approach.[14]

Lastly, we examined the doubly-robust DID approach proposed in Callaway and Sant'Anna,[15] developed to overcome the known bias in the classic TWFE model in the presence of treatment effect heterogeneity and staggered adoption.[15] CSA (like ASCM) aims to control for confounding in two stages: first by using weighting to ensure comparability between treated and control states within each enactment year cohort and then via additional covariate adjustment in the outcome model.

In this model, treated units are grouped into cohorts of units that started receiving treatment at the same time; untreated units are similarly aggregated. Then for each treated cohort and time period, a doubly-robust inference method is used to estimate the treatment effect for that group in that time period.[22]

This technique combines a probability-weighting approach and an outcome regression approach to provide a consistent estimator (in this case, for the average treatment effect on the treated) when either the propensity score model or the outcome regression model is correctly specified. This design-based approach is intended to minimize the impact of confounding bias. After estimates for each group and time period are calculated, the group-time treatment effects are aggregated to create an overall estimate of the treatment effect.

All of these methods involve the same core assumption: the difference in outcomes between the treated and untreated groups would remain constant in the absence of the policy intervention (with magnitude equal to that observed pre-policy). The models differ in how they control for confounders and baseline outcomes. The TWFE and AR models rely on parametric corrections (state and time-fixed effects for the TWFE model and lagged outcomes and time-fixed effects for the AR model); ASCM and CSA models use a design-based approach to improve comparability of treated and control states.

### 2.3 Simulation Details

Because it is impossible to test model assumptions in practice, we used a simulation study with a known data-generating process to assess the relative performance of these statistical models under scenarios with small, moderate, and large levels of confounding.

*Observed Data*

We derived data for the simulation from annual state-level data on outcomes and time-varying confounders from 1999-2016. The outcome of interest was the annual state-specific opioid mortality rate per 100,000 state residents, obtained from the 1999-2016 National Vital Statistics System Multiple Cause of Death mortality files. Consistent with other studies;[23-25] we identified opioid-related overdose deaths using *ICD10-CM*-external cause of injury codes X40-X44, X60-64, X85, and Y10-Y14. We used two time-varying confounders. The first was annual state-level unemployment rate,[26] frequently considered a confounder because unemployment rates are associated with multiple outcomes of interest as well as likelihood of policy adoption.[20] The second confounder was a time-varying function of either prior outcome levels or trends.

*Generating Simulation Data*

The simulation design built on prior work comparing statistical methods for evaluating the impact of state laws on total firearms deaths[6] and opioid-related mortality.[13]

*Generating Policy Enactment*

For each state and year (from 1999 to 2016), we generated a time-varying indicator $A_{it}$ to denote whether the policy was in effect. Once a policy was enacted, it remained in effect; thus, $A_{it}$ = 1 for all remaining years. For comparison states, $A_{it}$ = 0 for the entire study period. For policy states, we randomly generated month and year of policy enactment. In the first year of implementation, $A_{it}$ was coded as a fractional value between 0 and 1, indicating the percentage of the year in which the policy was in effect.

We considered the following *linear* and *nonlinear models* for generating policy enactment scenarios:

**Linear**: $Logit(\Pr(A_{it} = 1)) = b_0 + b_1 C_{it} + b_2 X_{it}$

**Nonlinear**: $Logit(\Pr(A_{it} = 1)) = b_0 + b_1 C_{it} + b_2 X_{it} + b_3 C_{it}^2 + b_4 X_{it}^2 + b_5 (C_{it} * X_{it})$

where $X_{it}$ is the state-level unemployment rate, and $C_{it}$ is a function of lagged outcome values. In other words, treatment assignment was associated with both the state unemployment rate and with prior state-level outcomes (as reflected in $C_{it}$).

For both the linear and nonlinear scenarios, we considered two approaches for computing $C_{it}$: *confounding by prior outcome levels*, in which $C_{it}$ equals the three-year prior moving average for each state in each year ($C_{it} = (Y^*_{i,t-3} + Y^*_{i,t-2} + Y^*_{i,t-1})/3$); and *confounding by prior outcome trends*, in which $C_{it}$ equals the change in the outcome level over the prior three years ($C_{it} = Y^*_{i,t-3} - Y^*_{i,t-1}$). We selected values of $b_0$ to ensure that a reasonable number of states (~ 22, ranging from 9 to 34 across iterations) would enact the policy over the course of the follow-up period.

*Generating Outcomes*

We generated outcome data for each state from 1999 to 2016 to ensure that confounders were associated with both the policy variable and the outcome and that the level of confounding could be modified:

**Linear**: $Y^*_{it} = Y_{it0} + a_1 C_{it} + a_2 X_{it} + \alpha A_{it}$

**Nonlinear**: $Y^*_{it} = Y_{it0} + a_1 C_{it} + a_2 X_{it} + a_3 C_{it}^2 + a_4 X_{it}^2 + a_5 (C_{it} * X_{it}) + \alpha A_{it}$

For untreated state-years ($Y_{it0}$), we set outcome values as equal to the actual observed state-specific, year-specific opioid overdose rates ($Y^{obs}_{it} = Y_{it0}$), plus additional terms from the two confounders ($C_{it}, X_{it}$). We generated outcomes for treated state-years ($Y_{it1}$) by augmenting the observed value $Y_{it0}$ with both the confounding terms and the policy effect of magnitude $\alpha$. We considered 2 policy effect conditions: null effect ($\alpha = 0$) and a non-null effect ($\alpha = -0.92$) that reflects approximately a 20% annual decrease in mortality. Results were highly similar, so results for the non-null runs are reported in eAppendix 1.

We anticipated that all simulation conditions would generate confounding, albeit of different magnitudes and forms. Our data-generating model used observed opioid-related data in which we can empirically demonstrate the presence of time-varying associations between unemployment level and prior outcome levels and trends, thereby supporting our selection of unemployment level as a confounder in this context[8] (see eAppendix 2 for details on our data).

We considered two approaches to augmenting this confounding (i.e., adding additional terms based on confounding by prior outcome trends and levels). We expected bias from both scenarios, although work by Zeldow & Hatfield suggests that conditions entailing confounding by outcome trends would generate greater bias than confounding by levels.[8] To ensure consistency across simulation scenarios, we selected values for $(b_1, b_2, b_3, b_4, b_5)$ and $(a_1, a_2, a_3, a_4, a_5)$ to produce small, moderate, and large levels of confounding. Our anchor for how much confounding was introduced for each set of parameter values was the standardized mean difference between simulated outcomes for policy states in which the policy was active compared to simulated outcomes from both states that never enacted the policy and policy states before policy enactment. We selected parameter values for each scenario that produced effect size differences of 0.15 (small confounding), 0.30 (medium confounding), and 0.45 (large confounding).

*Metrics for Assessing Relative Performance of Candidate Statistical Methods*

Performance metrics included bias, variance, root mean squared error, and coverage. For each combination of simulation parameters, we generated 5,000 simulated datasets.

(1) *Absolute Standardized Bias.* Bias assesses the average difference between the estimated effect and true effect over all simulations. To put bias on a similar scale across scenarios, we reported absolute standardized bias by dividing the estimated bias from each iteration by the estimated standard deviation of the outcome for each iteration. Thus, $AbsoluteBias_\alpha = \sum_{k=1}^{5000} \frac{|\hat{\alpha}_k - \alpha|}{5000 * sd(Y_{it}^*)}$.

(2) *Variance.* The empirical variance captures the spread of the estimated policy effects around the sample mean (in squared units) across the simulations. Thus, $Var_\alpha = \sum_{k=1}^{5000} \frac{(\hat{\alpha}_k - mean(\hat{\alpha}_k))^2}{5000}$

(3) *Root Mean Squared error (RMSE)*. RMSE is calculated by taking the square root of the sum of the mean squared errors (e.g., $\sqrt{\sum_{k=1}^{5000} (\hat{\alpha}_k - \alpha)^2 / 5000}$ ). RMSE quantifies error for a given model specification, considering both bias and variance.

(4) *Coverage*. The coverage probability is the proportion of time in which the 95% confidence interval for a given method contains the true value of the policy effect.

We conducted all simulations in R, using the "optic" package available in R. The package implements our simulation approach on user-provided outcome data. Code for reproducing these simulations for both a publicly available mortality data as well as user-provided data is provided in eAppendix 4. Users can use the package to explore additional confounding scenarios and candidate methods.

### 3. RESULTS

In our discussion, we highlight how bias, variance, and RMSE vary across simulation conditions in which magnitude of confounding; functional form of data generating processes (i.e., linear or nonlinear); and functional form of additional confounding terms (i.e., outcome trends or levels) vary. Table 1 describes the four methods studied, specifies how each addresses confounding, and presents key findings.

## 3.1 Absolute bias

Absolute bias increases for each method as the strength of confounding in the underlying data generating model increases (Figure 1). Bias is typically higher when there is confounding by prior outcome trends rather than levels and when the data generating model is nonlinear rather than linear. The only exception is the case of nonlinear confounding by prior outcome trends; here CSA provides superior performance.

For both the linear level and nonlinear level conditions, the TWFE and CSA methods are the most biased, with standardized bias ranging between 0.08 (small magnitude, linear level) and 0.28 (large magnitude, nonlinear level). Both AR and ASCM have standardized bias near 0 for virtually all magnitudes of confounding. Similar patterns appear in the linear trend condition.

In the nonlinear trend condition, the two best methods are ASCM and CSA, which have standardized bias below 0.25 for all magnitudes of confounding. Standardized bias for the AR and TWFE models can be sizeable (e.g., greater than 0.30 for large magnitude).

## 3.2 Variance

In each method, the empirical variance that captures the spread of the estimated policy effects around the sample mean (in squared units) across the simulation iterations is relatively constant across different magnitudes of confounding (Figure 2). In both linear and nonlinear settings, variance does not differ much between the levels and trends conditions,. The most notable impact on variance occurs when moving from the linear trend condition to the nonlinear trend condition: variances increase an average of 1.78-fold (ranging from 1.32 to 2.59). In contrast, comparing the linear level and nonlinear level conditions, variance is relatively similar for all methods except CSA in which variance is larger under the nonlinear level condition.

Performance regarding variance clearly delineates methods: ASCM has notably lower empirical variances than TWFE and CSA in all simulation conditions. The AR model also does well except for the nonlinear trend condition.

## 3.3 RMSE

Reflecting the findings regarding bias and variance, the RMSE tends to increase as the magnitude of confounding increases (Figure 3), driven primarily by bias increases (Figure 1). Additionally, RMSE is typically higher for trend conditions compared to level conditions and for nonlinear conditions compared to linear conditions. There are also meaningful RMSE differences across methods: AR and ASCM have lower RMSE than TWFE or CSA for all conditions except the nonlinear trend condition. For the linear level, linear trend, and nonlinear level conditions, ASCM has the lowest RMSE with AR a close second. For the nonlinear trend condition, ASCM outperforms all other methods on RMSE.

## 3.4 Coverage

Figure 4 shows nominal 95% confidence interval coverage rates for the four methods. Coverage decreases for all methods as the level of confounding increases: large confounding magnitude yields larger bias. Relatedly, we see worse coverage for the conditions in which the methods have greater bias (e.g., in trend conditions or nonlinear conditions). CSA has relatively stable coverage rates of around 80-85% across all simulation conditions; ASCM has coverage rates of 100%, due to its very high model-based variances when estimating treatments effects (results shown in eFigure 3a). Conversely, AR yields small (i.e., anti-conservative) model-based variances, which contribute to the very low coverage for AR in the nonlinear trend condition.

## 4. DISCUSSION

This novel simulation study examined how differing magnitudes of confounding affected the performance of TWFE, AR, ASCM, and CSA, statistical models commonly used in policy evaluation studies. No single method outperforms the others across simulation conditions. ASCM performs best when looking at RMSE, with low bias and lower empirical variance compared to other methods. However, coverage rates for ASCM exceeded the nominal 95% level, reflecting high model-based standard errors. Thus, a clear limitation of this method in practice is its wide (i.e., conservative) confidence intervals. The AR method also performs well, yielding low bias and more reasonable coverage rates in most simulation conditions; the exception is the nonlinear trend condition, in which the AR method yields large bias and low coverage rates (reflecting anti-conservative model-based standard errors).

Study findings highlight how performance of all methods declines as the magnitude of confounding increases in the prior outcome trends conditions; this is to be expected given that confounding by prior trends violates the key assumption that underlies each of these methods – namely that had the treated states not implemented the policy, their outcome trends would have evolved similarly to how outcomes evolved in the comparison states. In practice, the presence of confounding by prior outcome trends is highly likely in state policy evaluations, underscoring the need to use methods that can adequately address it. RMSE identified the ASCM as optimal for these cases, although coverage was poor (100% due to its high model-based standard errors). CSA performed best in the nonlinear trend condition, suggesting that this method also holds promise for applications in which confounding relationships are likely to be more complex.

It is important to consider the magnitude of confounding and its hypothetical influence on erroneous policy decisions. In our setting, we simulated the bias such that states with the poorest outcomes—in this case, opioid-related mortality—are more likely to enact a policy. Thus, in the null runs, all of the methods estimated a harmful policy effect. In practice, such study findings would be unlikely to influence adoption of a policy shown to be ineffective. However, adopting an ineffective policy is a critical concern because it diverts limited resources from policies that are effective.

In cases where a policy has a true protective effect (e.g., our non-null runs), the primary concern is that studies would erroneously identify either a null or a harmful effect, prompting policy makers to not enact or even to repeal a potentially helpful policy. Thus an inaccurate assessment could engender a missed opportunity to save lives (in the case of mortality outcomes) or to improve outcomes more generally for a population. It is important for applied studies to consider the implications of type I and II errors and to judge which may be more costly to society.

It is natural to consider using time-dependent confounding methods in this context, including longitudinal g-computation, inverse probability of treatment weighting (IPTW) estimators, augmented IPTW, or targeted maximum likelihood estimation.[27-34] Unfortunately, a challenge for using IPTW-based methods to evaluate state policies is the small sample size. Reliable IPTW cannot be attained because in any given year, only a few states will implement a new policy. ASCM uses a different type of weighting to create more comparable control and treated states and thus could be viewed as a proxy for these types of time-dependent confounding methods for state policy evaluations. We encourage future research to examine the potential of these methods for settings with larger sample sizes including education studies or examination of county-level policies.

Our findings must be considered alongside the study limitations. All the simulation settings had homogenous treatment effects that did not vary over time or location. Future work should explore the relative performance of methods in the presence of heterogenous treatment effects including those that change over time and/or location. CSA, for example, may perform relatively better in that case given that it is designed for such a setting. Other factors commonly present in opioid policy and other state policy evaluations but not included in our simulation design include: (1) the potential impact of averaging policy effects across states of different sizes/populations, (2) the impact of co-occurring policies[35] (e.g., prescription drug monitoring programs, naloxone distribution, cannabis policies), (3) heterogeneous policies across states (e.g., prescription drug monitoring programs can have different designs across states), and (4) heterogeneous implementation of policies within states, among others. Future work should explore how the models we examined and other methods might perform in such conditions. Finally, given the wide availability of more granular (e.g., county-level) data, future work should explore how well these models perform to study the heterogeneous effects of state-level policies across geographic areas and population subgroups.

Our assessment demonstrates that no single model outperforms the others across all simulation conditions. But a researcher's toolkit should include all methodological options. We simulated the performance of four models using opioid-related mortality data; however, our findings likely generalize to other state-level policy evaluations. Our simulation results and package in R can help researchers choose the most appropriate approach given available data, the outcomes being examined, and the nature of the policy decisions they hope to inform. Users can also explore additional confounding scenarios and candidate methods using the package.

We acknowledge that research evidence is only one factor that influences policy making; it also requires long-term and collaborative relationships between citizens, policymakers, health professionals and researchers that entails the exchange of rich information to help inform practice.[36-38]

**Uncategorized References**

28. Robins J. Causal inference from complex longitudinal data. In: *Latent Variable Modeling and Applications to Causality.* Springer; 1997:69-117.

29. Hernan MA, Robins JM. Estimating causal effects from epidemiological data. *J Epidemiol Community Health.* 2006;60(7):578-586.

30. Kurz CF. Augmented inverse probability weighting and the double robustness property. *Med Decis Making.* 2022;42(2):156-167.

31. Robins J. The analysis of randomized and non-randomized aids treatment trials using a new approach to causal inference in longitudinal studies. *Health Serv Res Methodol: A Focus on AIDS.* 1989:113-159.

32. Robins J, Hernan MA. Estimation of the causal effects of time-varying exposures. In: *Longitudinal Data Analysis.* New York, NY: Chapman and Hall/CRC Press; 2008:553-599.

33. Van der Laan M, Rose S. *Targeted Learning: Causal Inference for Observational and Experimental Data.* Vol 4. New York, NY: Springer; 2011.

34. Van der Laan M, Rubin D. Targeted maximum likelihood learning. *Int J Biostat.* 2006;2(1).

35. Griffin BA, Schuler MS, Pane J, et al. Methodological considerations for estimating policy effects in the context of co-occurring policies. *Health Serv Outcomes Res Methodol.* 2023;23(2):149-165.

36. Burstein P. The determinants of public policy: What matters and how much. *Policy Studies Journal.* 2020;48(1):87-110.

**Figure Legends**

**Figure 1.** Simulation results: Absolute standardized bias

**Figure 2.** Simulation results: Variance

**Figure 3.** Simulation results: Root mean squared error (RMSE)

**Figure 4.** Simulation results: Coverage

**Table 1. The four models examined, how each addresses confounding, and summary of performance**

| Method | How address confounding | Regression or design based | Susceptible scenarios | Key Findings |
| --- | --- | --- | --- | --- |
| Two-Way Fixed Effects (TWFE) | Covariate adjustment, state and time fixed effects | Regression | All | Higher bias, moderate variance, reasonable coverage |
| Autoregressive (AR) model | Covariate adjustment, including for lagged outcomes | Regression | Nonlinear trends | Otherwise, low bias & variance and reasonable coverage |
| Augmented Synthetic Control Method (ASCM) | Weighting based on pre-policy covariates and outcomes + covariate adjustment in outcome model | Design | Nonlinear trends | Minimal bias and low RMSE but high model based standard errors |
| Doubly robust difference-in-differences approach with multiple time periods from Callaway-Sant'Anna (CSA)[a] | Weighting based on pre-policy covariates and outcomes + covariate adjustment in outcome model | Design | Linear settings | Lower bias but high variance |

[a]We note that CSA is designed to estimate time-varying treatment effects; the focus of our simulation were scenarios with non-homogenous effects and thus we should be cautious when interpreting findings for CSA in this context.

**Figure 1. Simulation results: Absolute standardized bias**

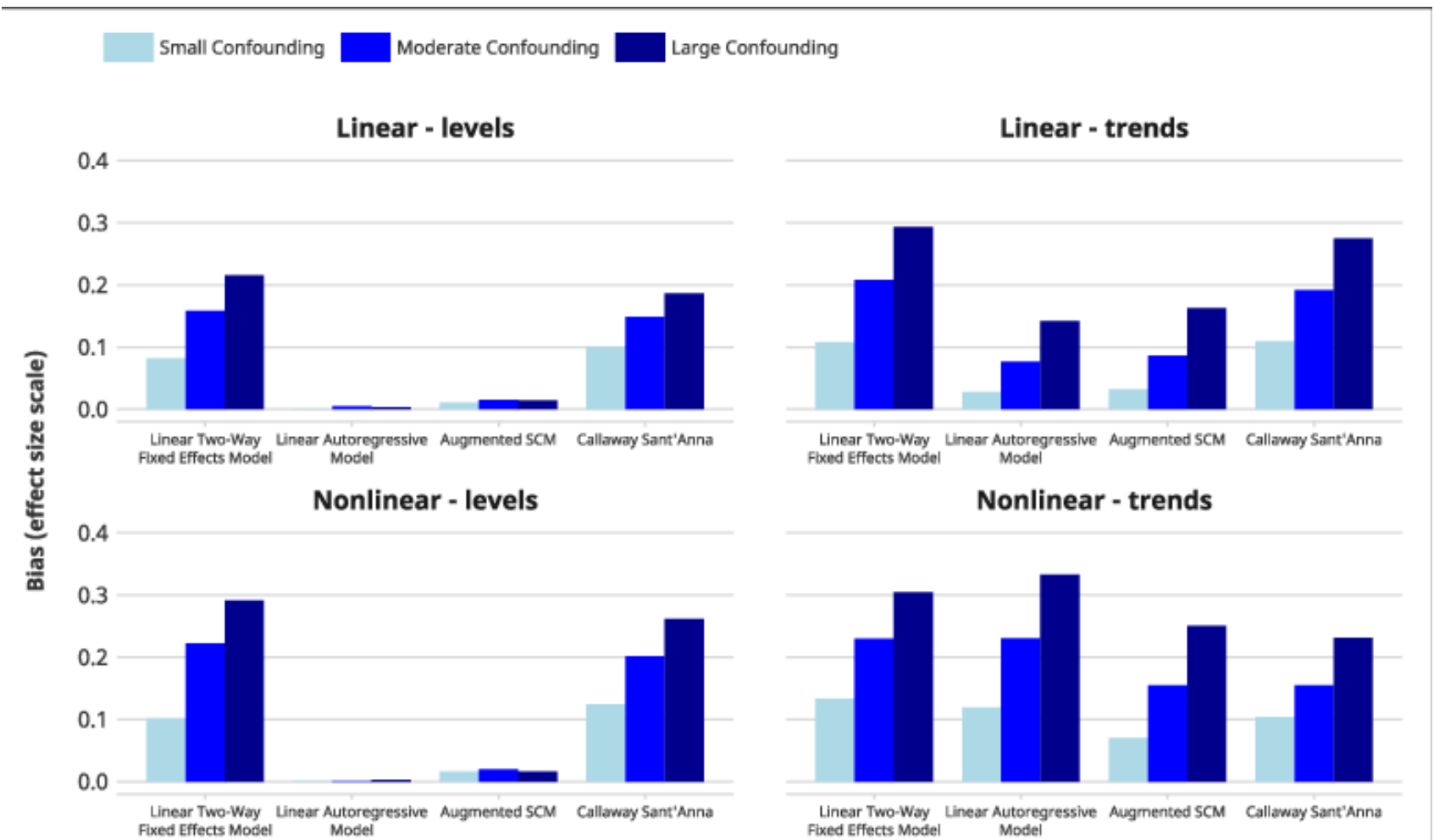

**Figure 2. Simulation results: Variance**

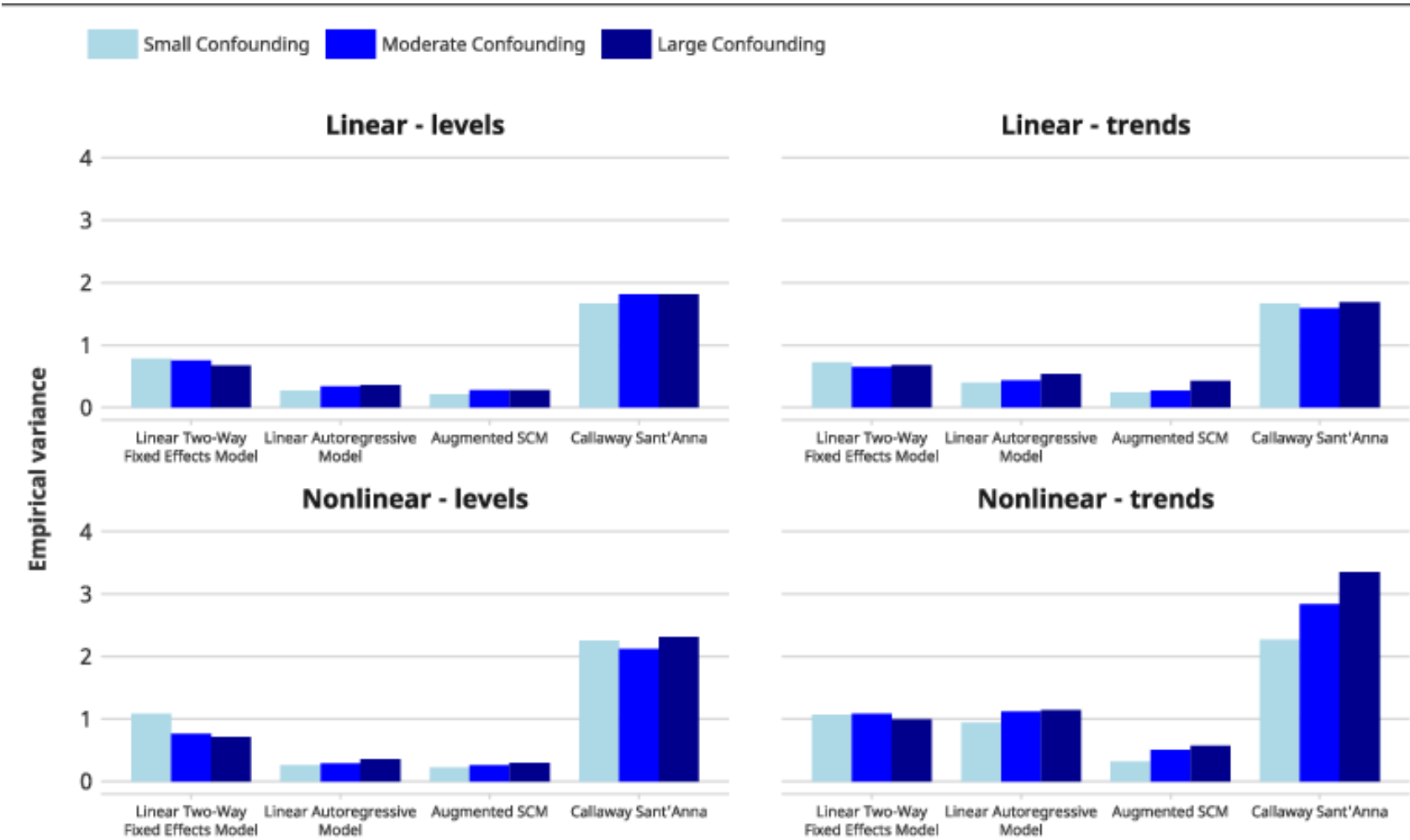

**Figure 3. Simulation results: Root mean squared error (RMSE)**

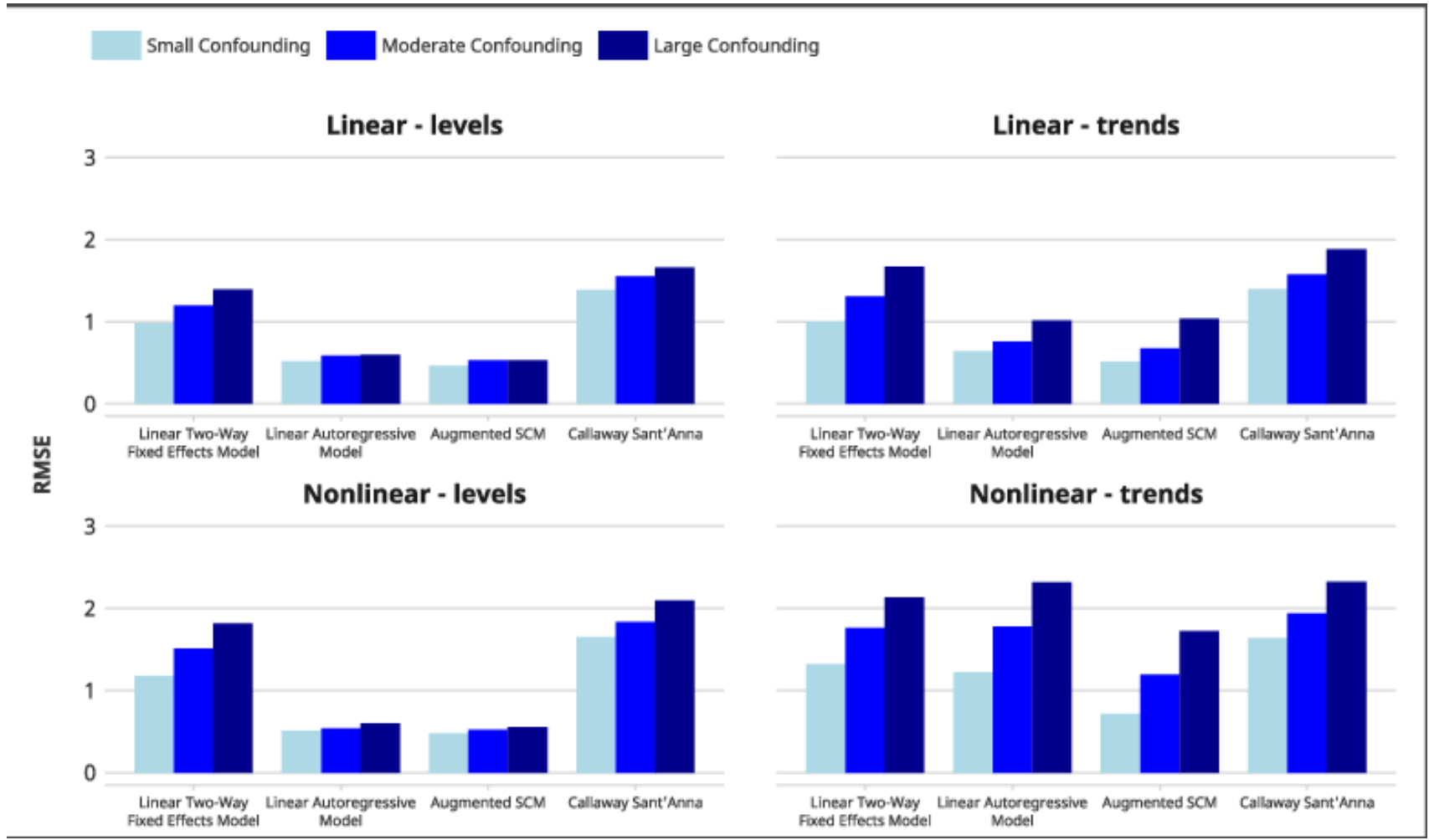

**Figure 4. Simulation results: 95% confidence interval coverage rates**

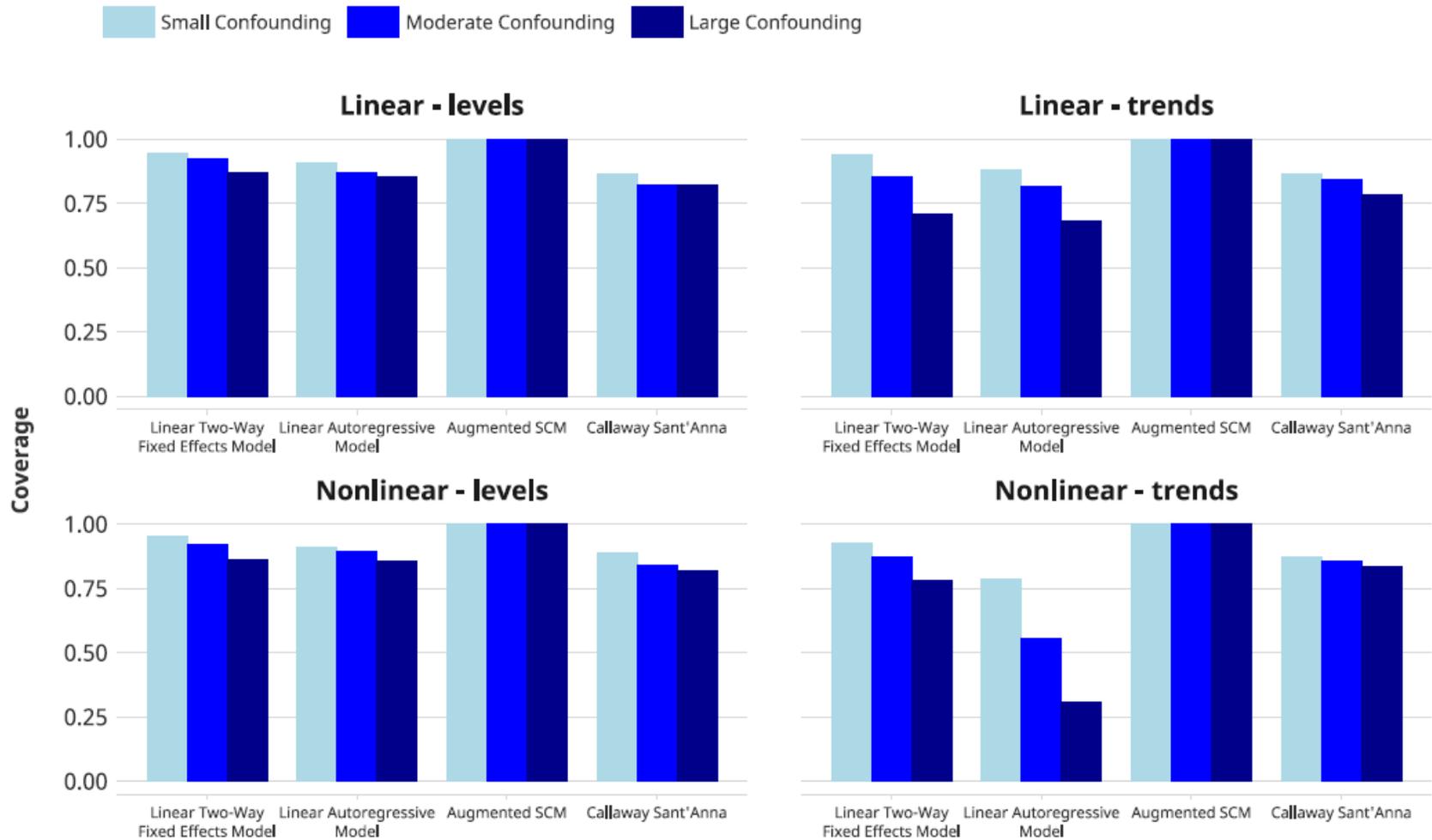

**eAppendix 1: Figures from the non-null runs**

**eFigure 1a. Simulation results for non-null run: Absolute standardized bias**

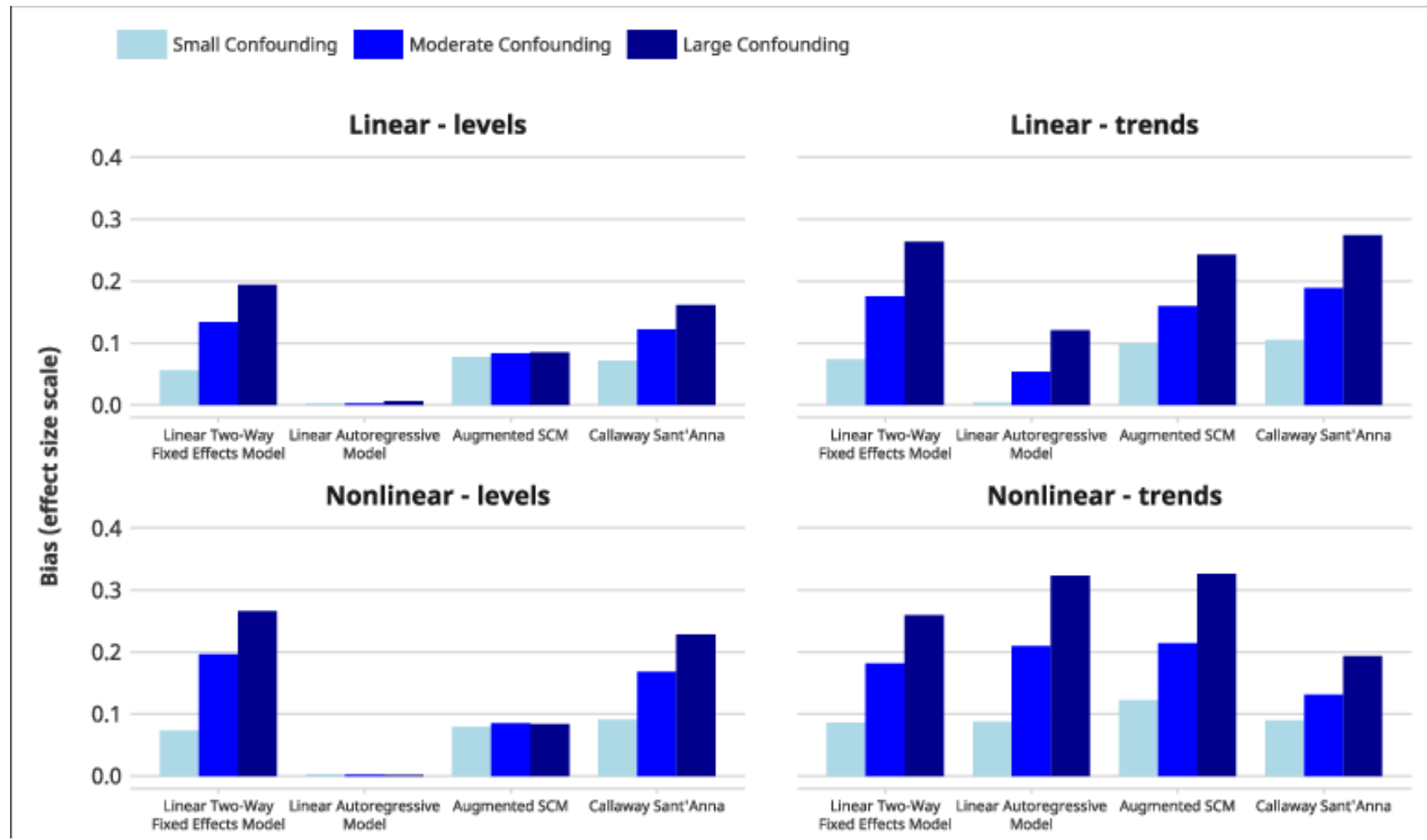

eFigure 1b. Simulation results for non-null run: Variance

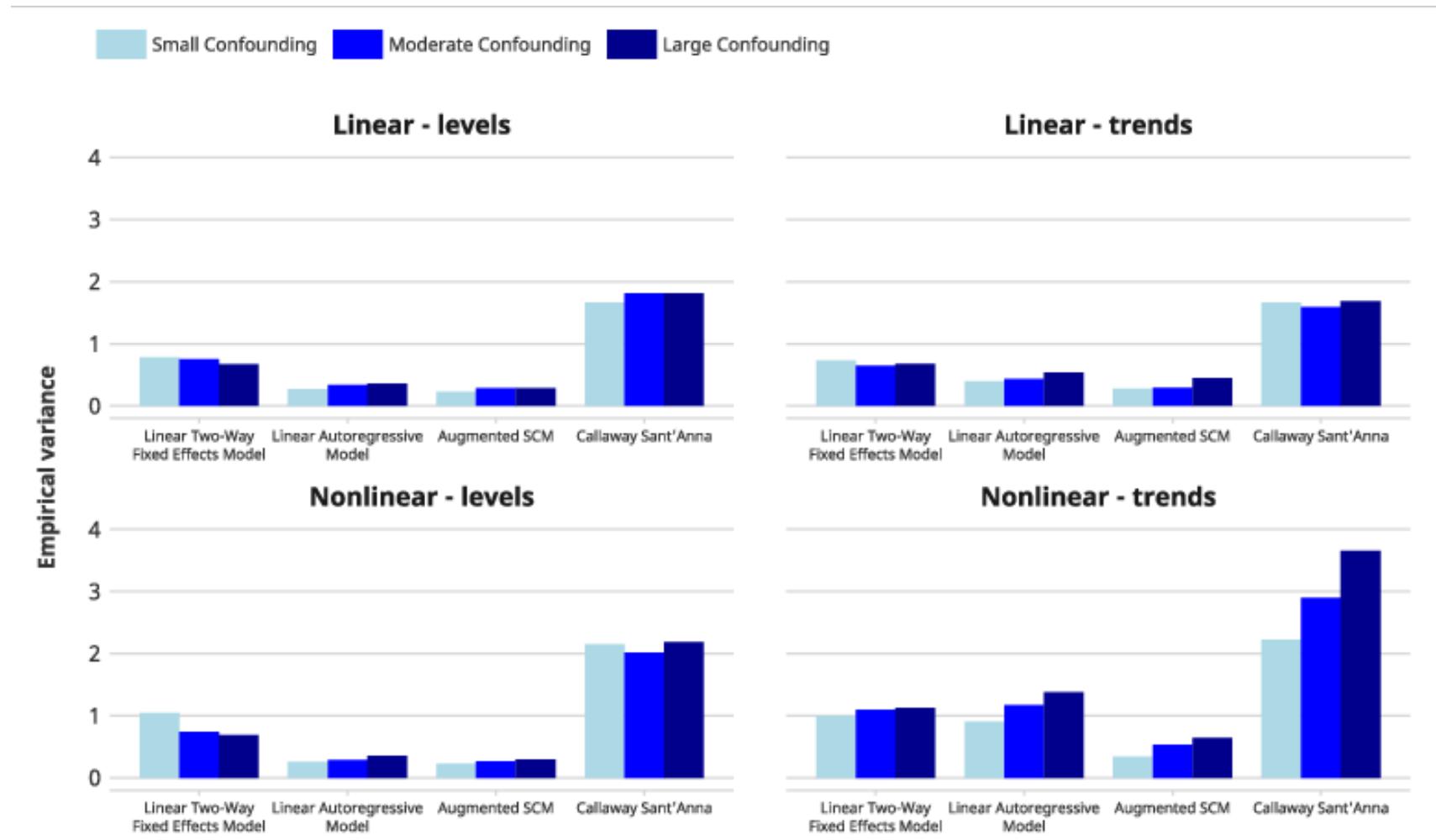

eFigure 1c. Simulation results for non-null run: Root mean squared error (RMSE)

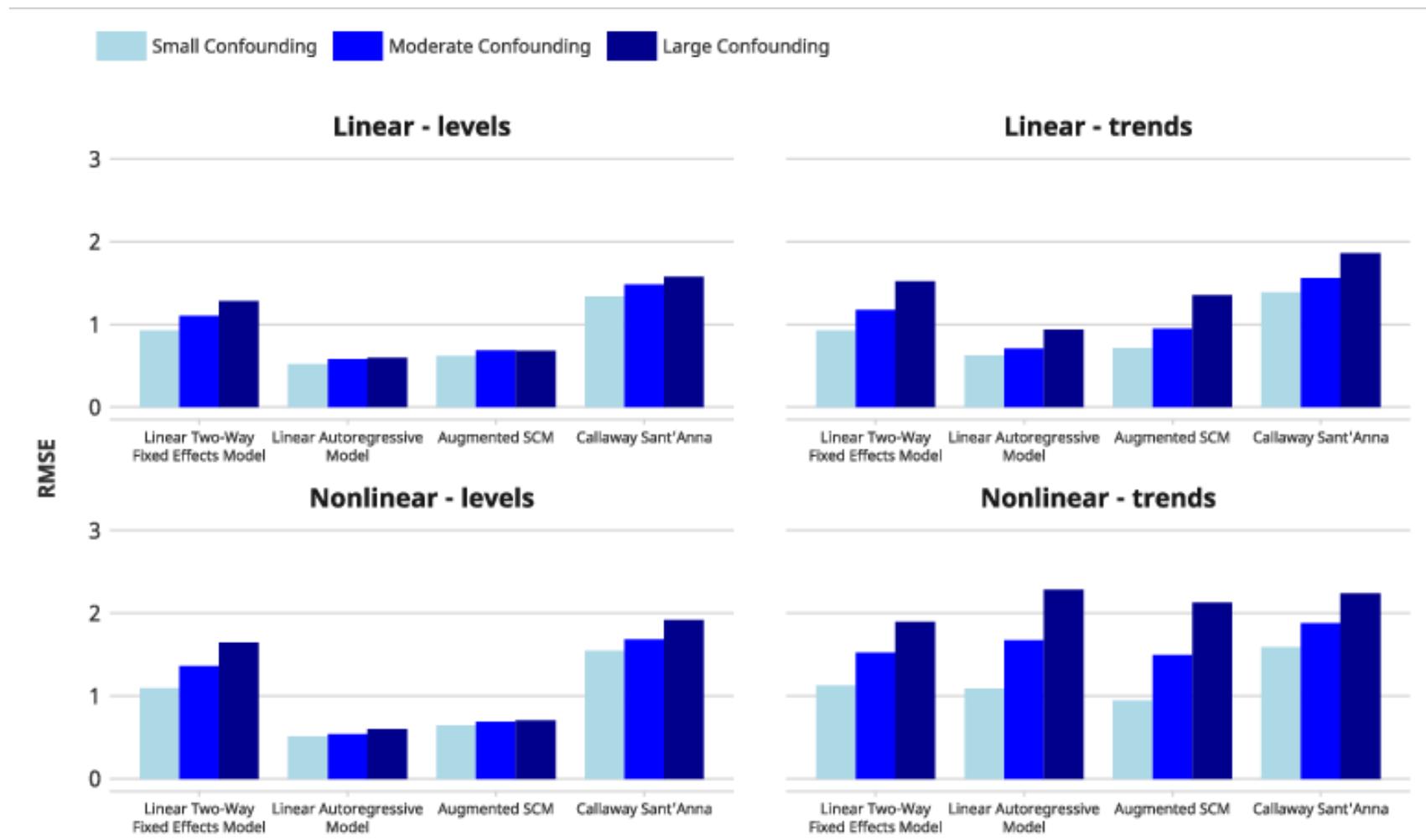

**eFigure 1d. Simulation results for non-null run: Coverage**

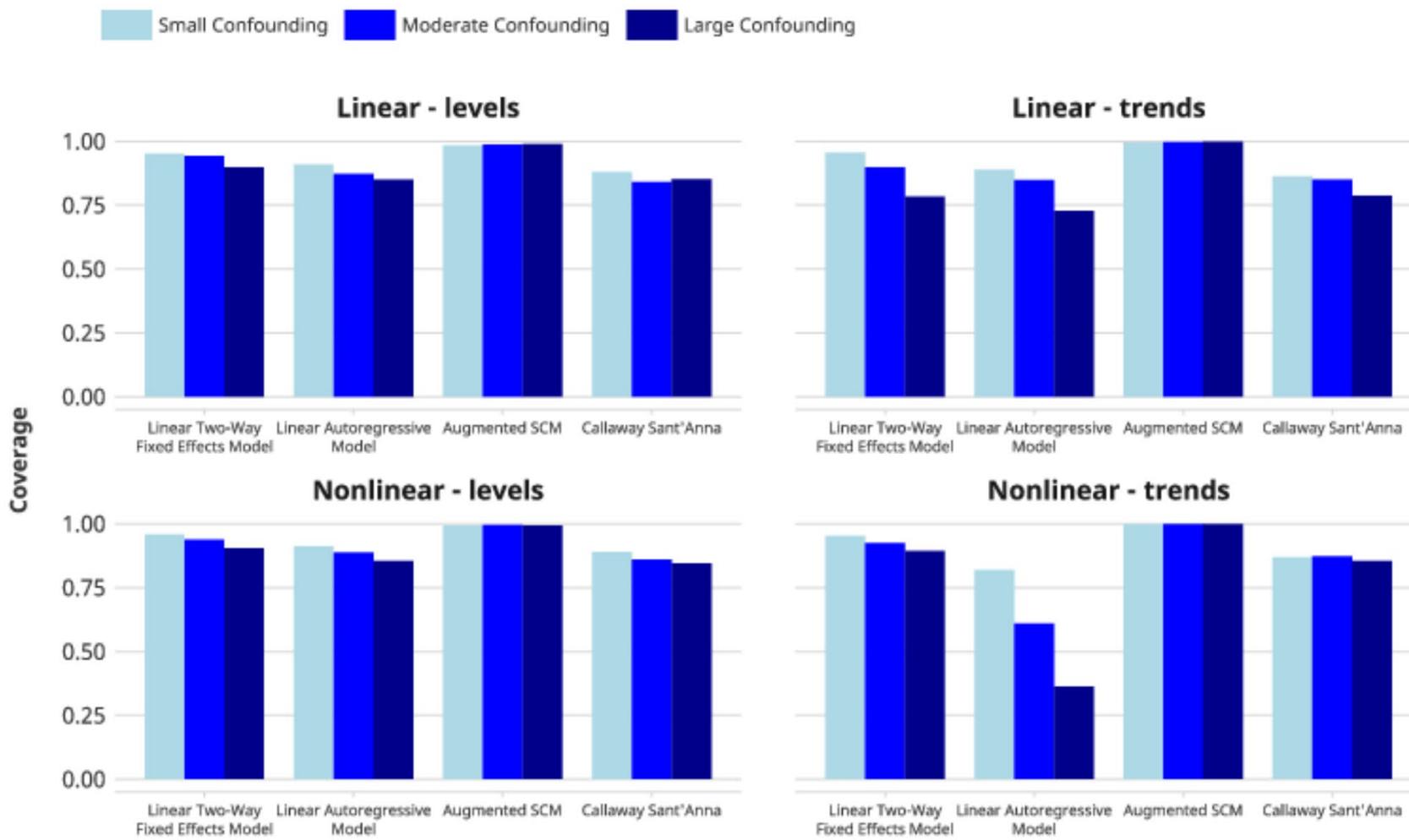

**eAppendix 2: Data descriptives**

**eFigure2a. Underlying time-varying relationship between unemployment rate and opioid mortality**

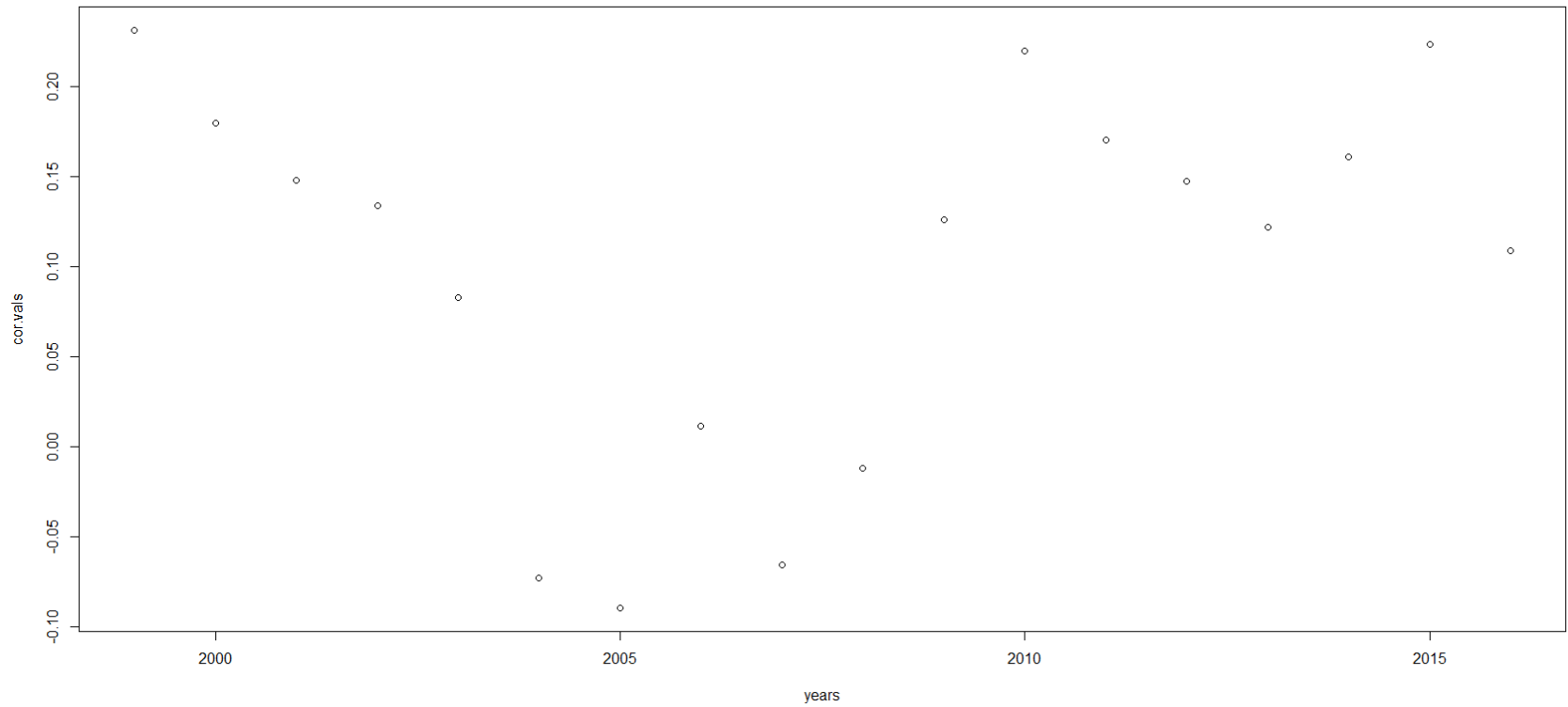

**eFigure2b. Trends over the follow-up for opioid mortality (solid line) and unemployment rates (dot-dashed line)**

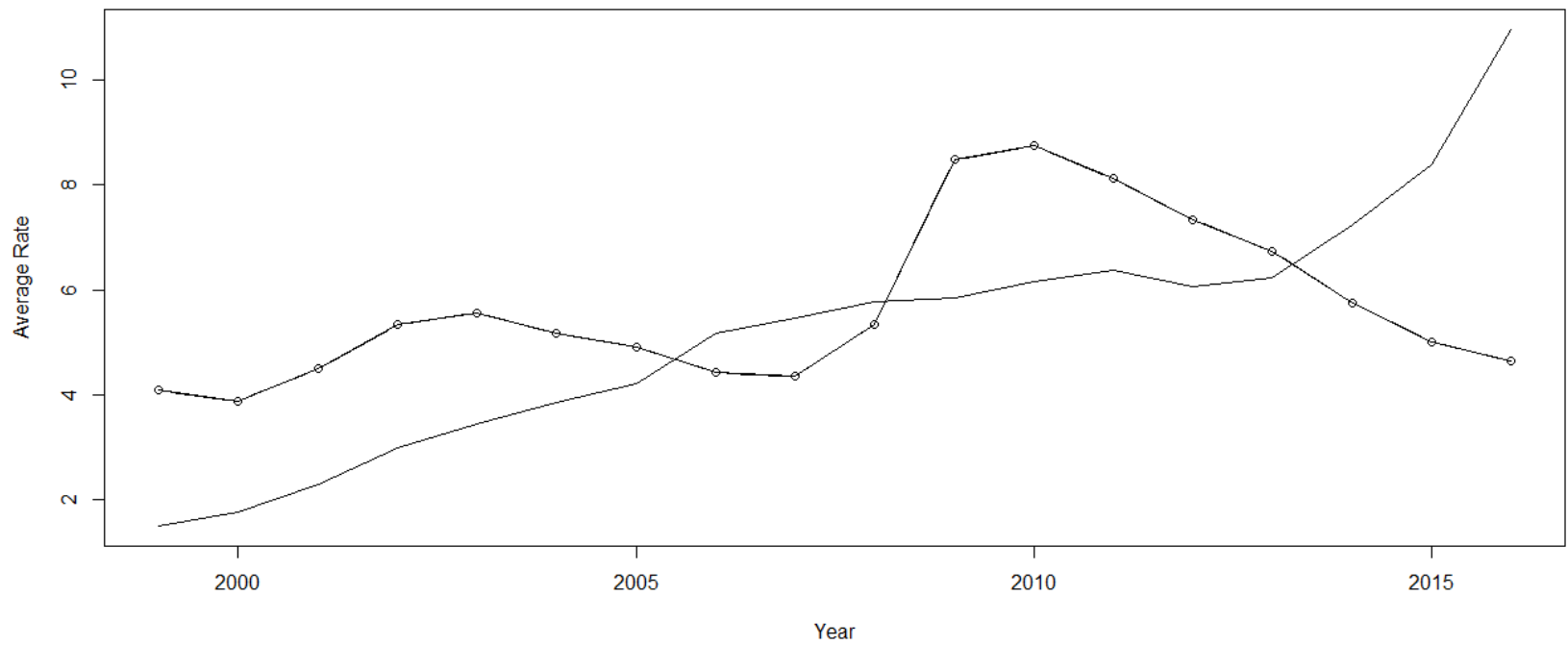

**eAppendix 3: Additional simulation results: Average model based variance and tabular results from all runs**

**eFigure3a. Simulation results from null runs: Average model-based variance**

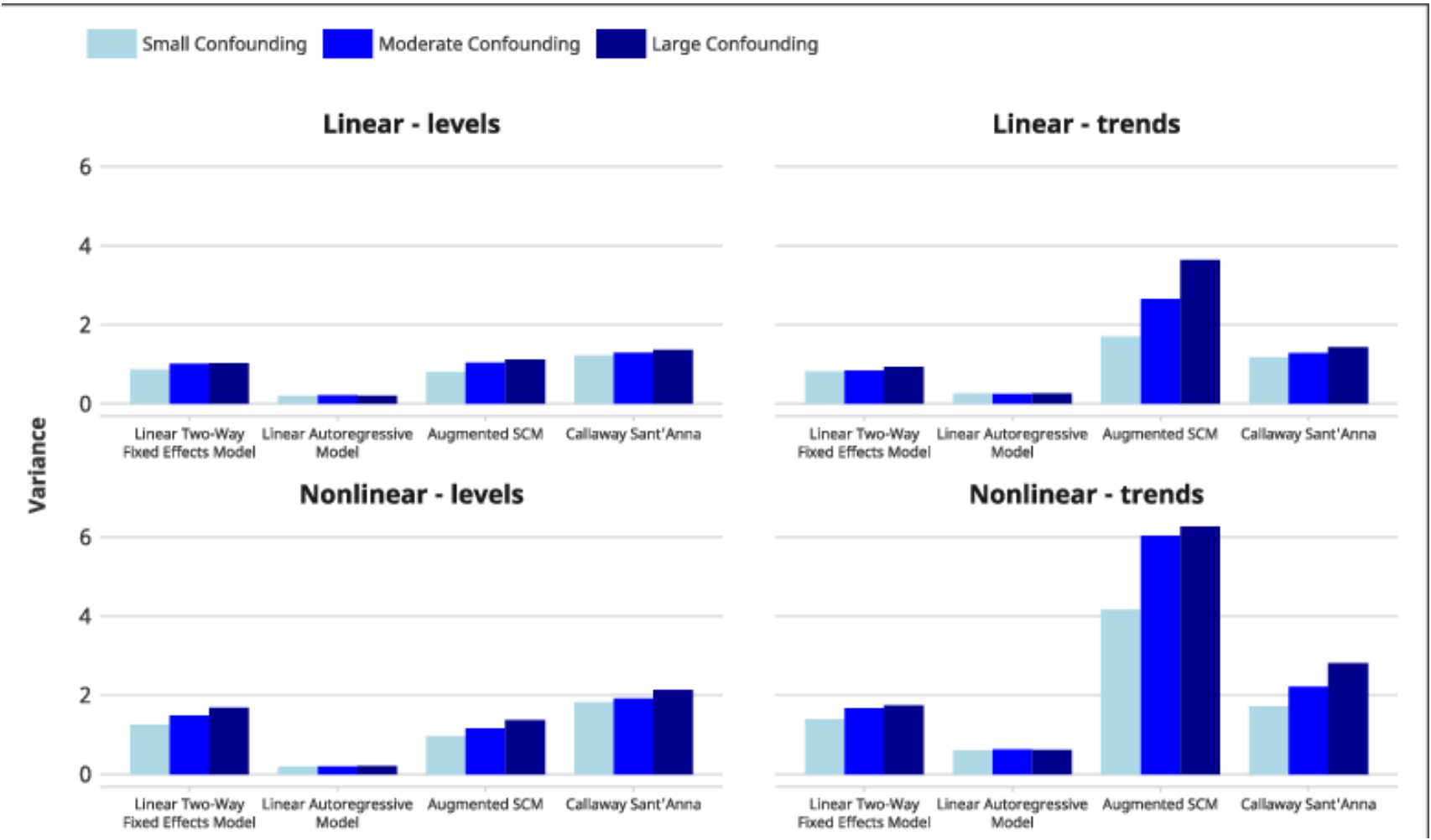

**eFigure3b. Simulation results from non-null runs: Average model-based variance**

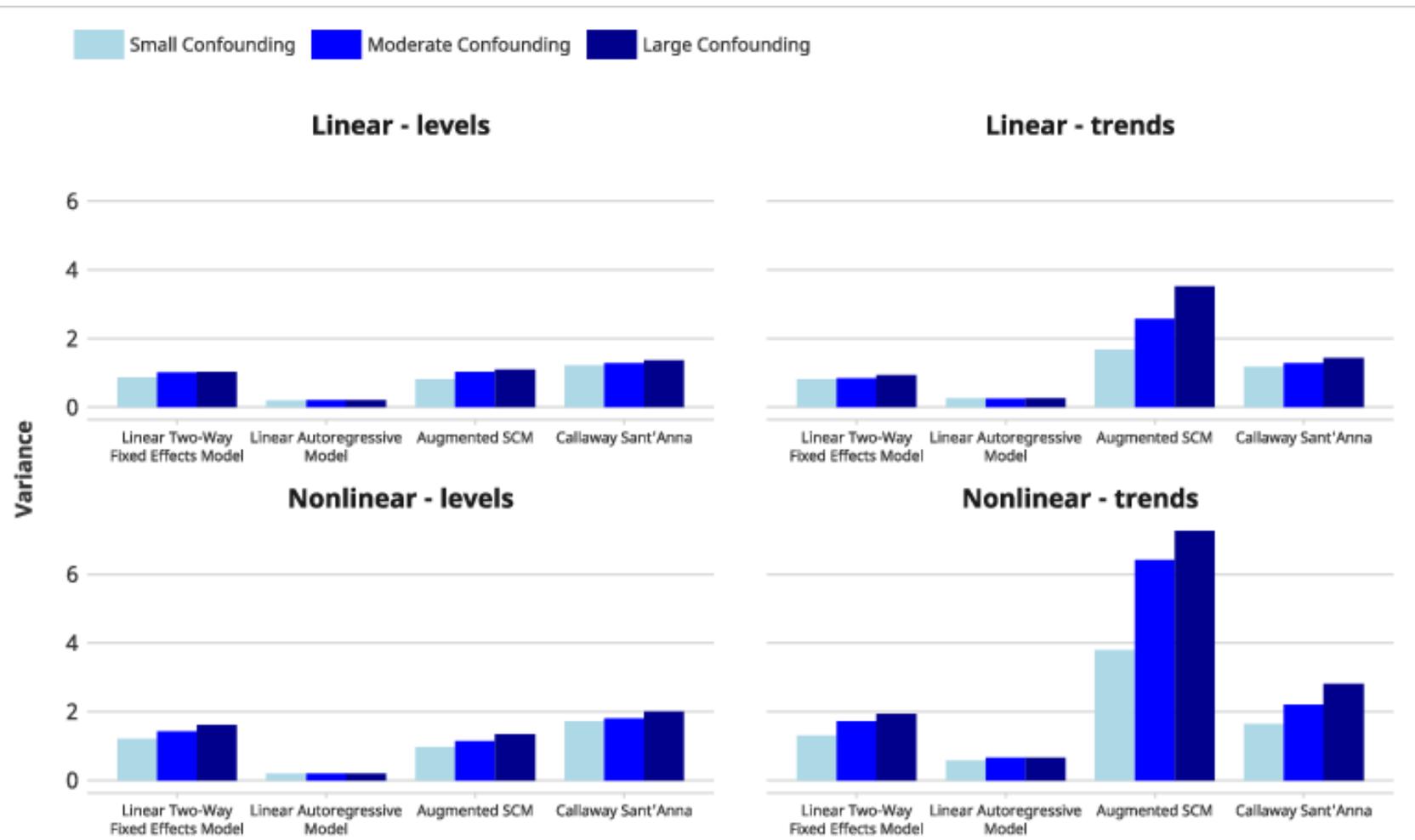

# eTable 3a – Detailed simulation results for the null runs for all scenarios and metrics

| Method | Confounding Type | Functional Form | Bias Size | Bias Effect Size Scale | MC Variance | RMSE | Coverage | Model-Based Variance |
|---|---|---|---|---|---|---|---|---|
| Augmented SCM | Levels | Linear | Small | 0.011 | 0.212 | 0.464 | 1.000 | 0.806 |
| Augmented SCM | Levels | Linear | Medium | 0.015 | 0.274 | 0.529 | 1.000 | 1.038 |
| Augmented SCM | Levels | Linear | Large | 0.014 | 0.276 | 0.531 | 1.000 | 1.112 |
| Callaway Sant'Anna | Levels | Linear | Small | 0.100 | 1.661 | 1.390 | 0.863 | 1.221 |
| Callaway Sant'Anna | Levels | Linear | Medium | 0.149 | 1.814 | 1.555 | 0.824 | 1.287 |
| Callaway Sant'Anna | Levels | Linear | Large | 0.187 | 1.811 | 1.661 | 0.824 | 1.360 |
| Linear Autoregressive Model | Levels | Linear | Small | 0.000 | 0.271 | 0.521 | 0.910 | 0.199 |
| Linear Autoregressive Model | Levels | Linear | Medium | 0.005 | 0.340 | 0.583 | 0.872 | 0.209 |
| Linear Autoregressive Model | Levels | Linear | Large | 0.003 | 0.358 | 0.598 | 0.852 | 0.201 |
| Linear Two-way Fixed Effects Model | Levels | Linear | Small | 0.082 | 0.779 | 0.980 | 0.947 | 0.868 |
| Linear Two-way Fixed Effects Model | Levels | Linear | Medium | 0.158 | 0.755 | 1.198 | 0.926 | 1.014 |
| Linear Two-way Fixed Effects Model | Levels | Linear | Large | 0.216 | 0.668 | 1.390 | 0.869 | 1.029 |
| Augmented SCM | Trends | Linear | Small | 0.032 | 0.239 | 0.514 | 1.000 | 1.698 |
| Augmented SCM | Trends | Linear | Medium | 0.086 | 0.271 | 0.674 | 0.998 | 2.652 |
| Augmented SCM | Trends | Linear | Large | 0.162 | 0.428 | 1.038 | 1.000 | 3.635 |
| Callaway Sant'Anna | Trends | Linear | Small | 0.109 | 1.668 | 1.401 | 0.864 | 1.171 |
| Callaway Sant'Anna | Trends | Linear | Medium | 0.192 | 1.588 | 1.578 | 0.845 | 1.278 |
| Callaway Sant'Anna | Trends | Linear | Large | 0.275 | 1.689 | 1.884 | 0.783 | 1.431 |
| Linear Autoregressive Model | Trends | Linear | Small | 0.028 | 0.392 | 0.642 | 0.882 | 0.257 |
| Linear Autoregressive Model | Trends | Linear | Medium | 0.077 | 0.432 | 0.759 | 0.814 | 0.250 |
| Linear Autoregressive Model | Trends | Linear | Large | 0.142 | 0.534 | 1.014 | 0.683 | 0.257 |
| Linear Two-way Fixed Effects Model | Trends | Linear | Small | 0.108 | 0.724 | 1.004 | 0.941 | 0.813 |
| Linear Two-way Fixed Effects Model | Trends | Linear | Medium | 0.208 | 0.647 | 1.307 | 0.855 | 0.843 |
| Linear Two-way Fixed Effects Model | Trends | Linear | Large | 0.293 | 0.678 | 1.670 | 0.711 | 0.934 |
| Augmented SCM | Levels | Nonlinear | Small | 0.016 | 0.217 | 0.475 | 1.000 | 0.964 |
| Augmented SCM | Levels | Nonlinear | Medium | 0.019 | 0.258 | 0.520 | 1.000 | 1.158 |
| Augmented SCM | Levels | Nonlinear | Large | 0.016 | 0.296 | 0.552 | 1.000 | 1.363 |

| Method | Confounding Type | Functional Form | Bias Size | Bias Effect Size Scale | MC Variance | RMSE | Coverage | Model-Based Variance |
|---|---|---|---|---|---|---|---|---|
| Callaway Sant'Anna | Levels | Nonlinear | Small | 0.124 | 2.254 | 1.651 | 0.887 | 1.820 |
| Callaway Sant'Anna | Levels | Nonlinear | Medium | 0.201 | 2.122 | 1.833 | 0.838 | 1.915 |
| Callaway Sant'Anna | Levels | Nonlinear | Large | 0.261 | 2.308 | 2.097 | 0.819 | 2.130 |
| Linear Autoregressive Model | Levels | Nonlinear | Small | 0.000 | 0.258 | 0.508 | 0.912 | 0.197 |
| Linear Autoregressive Model | Levels | Nonlinear | Medium | 0.000 | 0.290 | 0.539 | 0.894 | 0.196 |
| Linear Autoregressive Model | Levels | Nonlinear | Large | 0.003 | 0.357 | 0.598 | 0.856 | 0.199 |
| Linear Two-way Fixed Effects Model | Levels | Nonlinear | Small | 0.101 | 1.077 | 1.179 | 0.955 | 1.247 |
| Linear Two-way Fixed Effects Model | Levels | Nonlinear | Medium | 0.222 | 0.764 | 1.509 | 0.922 | 1.477 |
| Linear Two-way Fixed Effects Model | Levels | Nonlinear | Large | 0.291 | 0.707 | 1.817 | 0.864 | 1.673 |
| Augmented SCM | Trends | Nonlinear | Small | 0.070 | 0.319 | 0.714 | 1.000 | 4.160 |
| Augmented SCM | Trends | Nonlinear | Medium | 0.154 | 0.505 | 1.192 | 1.000 | 6.027 |
| Augmented SCM | Trends | Nonlinear | Large | 0.250 | 0.567 | 1.723 | 0.999 | 6.257 |
| Callaway Sant'Anna | Trends | Nonlinear | Small | 0.104 | 2.271 | 1.638 | 0.870 | 1.715 |
| Callaway Sant'Anna | Trends | Nonlinear | Medium | 0.155 | 2.840 | 1.940 | 0.859 | 2.210 |
| Callaway Sant'Anna | Trends | Nonlinear | Large | 0.231 | 3.350 | 2.325 | 0.837 | 2.810 |
| Linear Autoregressive Model | Trends | Nonlinear | Small | 0.119 | 0.942 | 1.220 | 0.787 | 0.600 |
| Linear Autoregressive Model | Trends | Nonlinear | Medium | 0.231 | 1.120 | 1.779 | 0.556 | 0.626 |
| Linear Autoregressive Model | Trends | Nonlinear | Large | 0.333 | 1.137 | 2.321 | 0.308 | 0.614 |
| Linear Two-way Fixed Effects Model | Trends | Nonlinear | Small | 0.133 | 1.063 | 1.320 | 0.928 | 1.389 |
| Linear Two-way Fixed Effects Model | Trends | Nonlinear | Medium | 0.230 | 1.084 | 1.764 | 0.870 | 1.667 |
| Linear Two-way Fixed Effects Model | Trends | Nonlinear | Large | 0.305 | 0.990 | 2.134 | 0.779 | 1.740 |

# eTable 3b – Detailed simulation results for the non-null runs for all scenarios and metrics

| Method | Confounding Type | Functional Form | Bias Size | Bias Effect Size Scale | MC Variance | RMSE | Coverage | Model-Based Variance |
|---|---|---|---|---|---|---|---|---|
| Augmented SCM | Levels | Linear | Small | 0.077 | 0.227 | 0.619 | 0.985 | 0.812 |
| Augmented SCM | Levels | Linear | Medium | 0.083 | 0.288 | 0.684 | 0.987 | 1.026 |
| Augmented SCM | Levels | Linear | Large | 0.084 | 0.281 | 0.682 | 0.99 | 1.094 |
| Callaway Sant'Anna | Levels | Linear | Small | 0.072 | 1.661 | 1.34 | 0.88 | 1.22 |
| Callaway Sant'Anna | Levels | Linear | Medium | 0.122 | 1.814 | 1.484 | 0.842 | 1.287 |
| Callaway Sant'Anna | Levels | Linear | Large | 0.161 | 1.811 | 1.576 | 0.852 | 1.36 |
| Linear Autoregressive Model | Levels | Linear | Small | 0.002 | 0.271 | 0.521 | 0.909 | 0.199 |
| Linear Autoregressive Model | Levels | Linear | Medium | 0.002 | 0.34 | 0.583 | 0.873 | 0.209 |
| Linear Autoregressive Model | Levels | Linear | Large | 0.006 | 0.358 | 0.599 | 0.85 | 0.201 |
| Linear Two-way Fixed Effects Model | Levels | Linear | Small | 0.056 | 0.779 | 0.927 | 0.952 | 0.868 |
| Linear Two-way Fixed Effects Model | Levels | Linear | Medium | 0.134 | 0.755 | 1.106 | 0.943 | 1.014 |
| Linear Two-way Fixed Effects Model | Levels | Linear | Large | 0.194 | 0.667 | 1.281 | 0.898 | 1.029 |
| Augmented SCM | Trends | Linear | Small | 0.099 | 0.276 | 0.713 | 0.995 | 1.675 |
| Augmented SCM | Trends | Linear | Medium | 0.16 | 0.294 | 0.95 | 0.997 | 2.576 |
| Augmented SCM | Trends | Linear | Large | 0.243 | 0.447 | 1.357 | 0.999 | 3.51 |
| Callaway Sant'Anna | Trends | Linear | Small | 0.105 | 1.668 | 1.389 | 0.864 | 1.171 |
| Callaway Sant'Anna | Trends | Linear | Medium | 0.189 | 1.589 | 1.56 | 0.851 | 1.278 |
| Callaway Sant'Anna | Trends | Linear | Large | 0.274 | 1.69 | 1.862 | 0.788 | 1.431 |
| Linear Autoregressive Model | Trends | Linear | Small | 0.004 | 0.393 | 0.627 | 0.89 | 0.259 |
| Linear Autoregressive Model | Trends | Linear | Medium | 0.053 | 0.432 | 0.707 | 0.849 | 0.251 |
| Linear Autoregressive Model | Trends | Linear | Large | 0.12 | 0.535 | 0.936 | 0.729 | 0.258 |
| Linear Two-way Fixed Effects Model | Trends | Linear | Small | 0.074 | 0.726 | 0.926 | 0.956 | 0.813 |
| Linear Two-way Fixed Effects Model | Trends | Linear | Medium | 0.175 | 0.649 | 1.175 | 0.898 | 0.843 |
| Linear Two-way Fixed Effects Model | Trends | Linear | Large | 0.263 | 0.68 | 1.523 | 0.784 | 0.935 |
| Augmented SCM | Levels | Nonlinear | Small | 0.079 | 0.229 | 0.643 | 0.994 | 0.965 |
| Augmented SCM | Levels | Nonlinear | Medium | 0.085 | 0.261 | 0.686 | 0.996 | 1.138 |
| Augmented SCM | Levels | Nonlinear | Large | 0.083 | 0.295 | 0.703 | 0.994 | 1.336 |

| Method | Confounding Type | Functional Form | Bias Size | Bias Effect Size Scale | MC Variance | RMSE | Coverage | Model-Based Variance |
|---|---|---|---|---|---|---|---|---|
| Callaway Sant'Anna | Levels | Nonlinear | Small | 0.091 | 2.151 | 1.547 | 0.891 | 1.722 |
| Callaway Sant'Anna | Levels | Nonlinear | Medium | 0.168 | 2.019 | 1.685 | 0.861 | 1.798 |
| Callaway Sant'Anna | Levels | Nonlinear | Large | 0.228 | 2.187 | 1.918 | 0.846 | 1.993 |
| Linear Autoregressive Model | Levels | Nonlinear | Small | 0.002 | 0.258 | 0.508 | 0.912 | 0.197 |
| Linear Autoregressive Model | Levels | Nonlinear | Medium | 0.002 | 0.29 | 0.539 | 0.889 | 0.196 |
| Linear Autoregressive Model | Levels | Nonlinear | Large | 0.001 | 0.356 | 0.597 | 0.854 | 0.199 |
| Linear Two-way Fixed Effects Model | Levels | Nonlinear | Small | 0.073 | 1.046 | 1.097 | 0.958 | 1.206 |
| Linear Two-way Fixed Effects Model | Levels | Nonlinear | Medium | 0.196 | 0.741 | 1.361 | 0.939 | 1.42 |
| Linear Two-way Fixed Effects Model | Levels | Nonlinear | Large | 0.266 | 0.686 | 1.647 | 0.904 | 1.604 |
| Augmented SCM | Trends | Nonlinear | Small | 0.122 | 0.339 | 0.946 | 1 | 3.783 |
| Augmented SCM | Trends | Nonlinear | Medium | 0.214 | 0.533 | 1.493 | 1 | 6.418 |
| Augmented SCM | Trends | Nonlinear | Large | 0.326 | 0.641 | 2.132 | 0.999 | 7.253 |
| Callaway Sant'Anna | Trends | Nonlinear | Small | 0.09 | 2.226 | 1.59 | 0.869 | 1.634 |
| Callaway Sant'Anna | Trends | Nonlinear | Medium | 0.131 | 2.895 | 1.88 | 0.873 | 2.196 |
| Callaway Sant'Anna | Trends | Nonlinear | Large | 0.193 | 3.654 | 2.242 | 0.855 | 2.81 |
| Linear Autoregressive Model | Trends | Nonlinear | Small | 0.087 | 0.901 | 1.09 | 0.819 | 0.58 |
| Linear Autoregressive Model | Trends | Nonlinear | Medium | 0.21 | 1.169 | 1.672 | 0.61 | 0.654 |
| Linear Autoregressive Model | Trends | Nonlinear | Large | 0.324 | 1.378 | 2.283 | 0.363 | 0.66 |
| Linear Two-way Fixed Effects Model | Trends | Nonlinear | Small | 0.086 | 0.999 | 1.129 | 0.954 | 1.295 |
| Linear Two-way Fixed Effects Model | Trends | Nonlinear | Medium | 0.182 | 1.093 | 1.522 | 0.926 | 1.713 |
| Linear Two-way Fixed Effects Model | Trends | Nonlinear | Large | 0.26 | 1.126 | 1.896 | 0.894 | 1.937 |

# eAppendix 4: Code for reproduction of runs on publicly available data

```
#---------------------------------------------------------------------------#
# OPTIC R Package Code Repository
# Copyright (C) 2023 by The RAND Corporation
# See README.md for information on usage and licensing
#---------------------------------------------------------------------------#

# Running an example of the confounding method for reproduction of paper

library(optic)
library(glmnet)
library(dplyr)
library(augsynth)
library(did)
library(future)
library(future.apply)
library(glue)
library(arrow)

data(overdoses)

linear0 <- 0
linear5 <- .05*mean(overdoses$crude.rate, na.rm=T)
linear15 <- .15*mean(overdoses$crude.rate, na.rm=T)
linear25 <- .25*mean(overdoses$crude.rate, na.rm=T)

# create optic_model object:
fixedeff_linear <- optic_model(
  name="fixedeff_linear",
```

```r
    type="reg",
    call="lm",
    formula=crude.rate ~ treatment_level + unemploymentrate + as.factor(year) + as.factor(state),
    weights=as.name("population"),
    se_adjust=c("none", "cluster")
)

lm_ar <- optic_model(
    name = "auto_regressive_linear",
    type = "autoreg",
    call = "lm",
    formula = crude.rate ~ unemploymentrate + as.factor(year) + treatment_change,
    se_adjust=c("none", "cluster")
)

multisynth_model <- list(
    name="multisynth",
    type="multisynth",
    model_call="multisynth",
    model_formula=crude.rate ~ treatment_level | unemploymentrate,
    model_args=list(unit=as.name("state"), time=as.name("year"), fixedeff=TRUE,
            form=crude.rate ~ treatment_level | unemploymentrate),
    se_adjust="none"
)

did_model <- list(
    name="did",
    type="did",
    model_call="att_gt",
```

```r
    model_formula= ~ unemploymentrate,
    model_args=list(yname="crude.rate", tname="year", idname="state",
             gname="treatment_year", xformla = formula("~ unemploymentrate")),
    se_adjust=c("none")
)

# Creating bias vals object - these values are used in confounding paper sim
bias_vals <- list(
  linear = list(
    level = list(
      small=c(b0=-3.9, b1=0.06, b2=0.06, b3=0, b4=0, b5=0,
          a1=0.2, a2=0.05, a3=0, a4=0, a5=0),
      medium=c(b0=-4.3, b1=0.11, b2=0.07, b3=0, b4=0, b5=0,
           a1=0.2, a2=0.05, a3=0, a4=0, a5=0),
      large=c(b0=-4.7, b1=0.16, b2=0.1, b3=0, b4=0, b5=0,
          a1=0.2, a2=0.05, a3=0, a4=0, a5=0),
      none = c(b0=-4.8, b1=0, b2=0, b3=0, b4=0, b5=0,
           a1=0, a2=0, a3=0, a4=0, a5=0)),
    trend = list(
      small=c(b0=-3.7, b1=0.15, b2=0.05, b3=0, b4=0, b5=0,
          a1=0.5, a2=0.11, a3=0, a4=0, a5=0),
      medium=c(b0=-4.5, b1=0.26, b2=0.16, b3=0, b4=0, b5=0,
           a1=0.5, a2=0.11, a3=0, a4=0, a5=0),
      large=c(b0=-5.1, b1=0.37, b2=0.22, b3=0, b4=0, b5=0,
          a1=0.5, a2=0.11, a3=0, a4=0, a5=0),
      none = c(b0=-5, b1=0, b2=0, b3=0, b4=0, b5=0,
           a1=0, a2=0, a3=0, a4=0, a5=0))),
  nonlinear = list(
    level = list(
```

```r
      small=c(b0=-3.8, b1=0.05, b2=0.05, b3=0.0003, b4=0.0003, b5=0.000003,
           a1=0.01, a2=0.01, a3=0.01, a4=0.01, a5=0.001),
      medium=c(b0=-4, b1=0.05, b2=0.05, b3=0.003, b4=0.003, b5=0.00003,
            a1=0.01, a2=0.01, a3=0.01, a4=0.01, a5=0.001),
      large=c(b0=-4.2, b1=0.05, b2=0.05, b3=0.0055, b4=0.0055, b5=0.000055,
           a1=0.01, a2=0.01, a3=0.01, a4=0.01, a5=0.001)),
    trend = list(
      small=c(b0=-3.6, b1=0.05, b2=0.001, b3=0.005, b4=0.008, b5=0.005,
           a1=0.1, a2=0.05, a3=0.1, a4=0.01, a5=0.01),
      medium=c(b0=-4.4, b1=0.05, b2=0.02, b3=0.018, b4=0.015, b5=0.015,
            a1=0.1, a2=0.05, a3=0.1, a4=0.01, a5=0.01),
      large=c(b0=-5.1, b1=0.05, b2=0.03, b3=0.025, b4=0.025, b5=0.025,
           a1=0.1, a2=0.05, a3=0.1, a4=0.01, a5=0.01)
    )
  )
)

set.seed(894539)

# Filtering SOuth Dakota and North Dakota because they have NA's in their
# crude rate variable

data <- overdoses %>%
  dplyr::filter(!(state %in% c("South Dakota", "North Dakota")))

linear_fe_config <- optic_simulation(
  x=data,
  models=list(fixedeff_linear, lm_ar,multisynth_model),
  iters=5,
```

```
    method = "confounding",
    globals=list(bias_vals=bias_vals),
    unit_var="state",
    treat_var="state",
    time_var="year",
    conf_var = "unemploymentrate",
    effect_magnitude=list(linear0),
    n_units= c(5),
    effect_direction=c("null"), #update and run with nonnull TE below once in good place
    policy_speed=list("instant"),
    n_implementation_periods=c(0),
    prior_control=c("trend", "level"),
    bias_type=c("linear","nonlinear"),
    bias_size=c("small","medium","large")
)

linear_results <- dispatch_simulations(
    linear_fe_config,
    use_future=T,
    seed=9782,
    verbose=2,
    future.globals=c("cluster_adjust_se"),
    future.packages=c("MASS", "dplyr", "optic","augsynth","did")
)

linear_results_df <- do.call(rbind, linear_results)

linear_results_df
```

```r
write.table(linear_results_df,"confounding_runs_casestudy_nullruns.csv",sep=",")
```